\documentclass{elsart}

\usepackage{epsfig}

\usepackage{amssymb}

\usepackage{graphicx} 
\usepackage{amsmath}

\begin{document}

\begin{frontmatter}



\title{A Coupled Equations Model for Epitaxial Growth on Textured Surfaces}


\author[UBC,cor1]{A.~Ballestad},
\ead{anders@physics.ubc.ca}
\ead[url]{http://www.physics.ubc.ca/mbelab}
\author[UBC,UBC2]{T.~Tiedje},
\author[UBC]{J.~H.~Schmid},
\author[NZ]{B.~J.~Ruck} and
\author[OCP]{M.~Adamcyk} 
\address[UBC]{Department of Physics and Astronomy, University of
  British Columbia, Vancouver, BC V6T 1Z1, Canada}
\address[UBC2]{Also Department of Electrical and Computer Engineering, University of
  British Columbia, Vancouver, BC V6T 1Z4, Canada}
\address[NZ]{School of Chemical and Physical Sciences,
  Victoria University of Wellington, New Zealand}
\address[OCP]{Optical Communication Products Inc.,
  Bloomfield, CO, USA}
\corauth[cor1]{Corresponding author. Phone: +1-604-822-5425, fax: +1-604-822-4750}

\begin{abstract}
  We have developed a coupled equations continuum model that explains
  the complex surface shapes observed in epitaxial regrowth on micron
  scale gratings.  This model describes the dependence of the surface
  morphology on film thickness and growth temperature in terms of a
  few simple atomic scale processes including adatom diffusion,
  step-edge attachment and detachment, and a net downhill migration of
  surface adatoms. The continuum model reduces to the linear part of
  the Kardar-Parisi-Zhang equation with a flux dependent smoothing
  coefficient in the long wavelength limit.
\end{abstract}

\begin{keyword}
Theory and models of crystal growth \sep 
Physics of crystal growth \sep
GaAs surface morphology \sep
GaAs homoepitaxy
\PACS 81.10.Aj \sep 81.15.Aa \sep 68.55.-a
\end{keyword}
\end{frontmatter}

\section{Introduction}
\label{intro}

The problem of the time evolution of the shape of crystal surfaces has
a long history dating back to Mullins and Herring who considered
relaxation during annealing above the roughening
temperature~\cite{Pimpinelli}. More recently, shape relaxation below
the roughening temperature has been studied
extensively~\cite{GreenBook,Erlebacher,Shenoy}. Below the roughening
temperature the problem is complicated by the need to keep track of
the dynamics of atomic steps and the fact that the surface free energy
of crystal facets is singular. Biasiol {\it et al.}~\cite{Biasiol}
have extended the theory of shape relaxation below the roughening
temperature to include the effects of atom deposition, and use this
theory to explain the self limiting V-grooves observed in
organo-metallic chemical vapor deposition (OMCVD) growth on corrugated
GaAs substrates.  In this paper we present a new continuum model which
we use to interpret measurements of the shape of corrugated GaAs (100)
surfaces under growth conditions which do not produce faceting. Facets
are not present in our experiments due to atomic scale roughness
associated with atom deposition in the island growth mode, and the
fact that the surface topography is sufficiently weak that the surface
slope does not reach the low energy [111] facets.  We show that this
model reproduces the surface morphology that develops during molecular
beam epitaxy (MBE) regrowth on 1D surface gratings.

\section{Conventional Modeling of Weak Surface Texture}
\label{cont}

The evolution of long wavelength surface structures during GaAs
homoepitaxy can be described by the Kardar-Parisi-Zhang (KPZ)
equation~\cite{Pimpinelli,Ballestad2}: $\partial h/\partial t = \nu
\nabla^2h + \frac{\lambda}{2}(\nabla h)^2 + F+ \eta(x,t)$.  The
coefficients in this equation are constants characterizing the
microscopic atomic processes.  The source term $\eta(x,t)$ simulates
the random arrival of atoms at an average rate $F$.  According to this
equation, a textured starting surface will develop parabolic mounds
that smooth with time separated by V-shaped valleys.  Recent
experimental work~\cite{Ballestad} has shown that the KPZ equation
provides an accurate description of the morphology of epitaxially
grown GaAs layers for surfaces with local slopes $\lesssim 3^{\circ}$.
The agreement with this simple continuum model suggests that the
anomalous effects associated with the singular free energy of crystal
facets are not important for the growth conditions in question.

In the case of GaAs molecular beam epitaxy (MBE) growth in which there
is no re-evaporation, the simplest explanation for the linear term in
the KPZ equation is that it is due to an inverse Ehrlich-Schwoebel
(ES) effect~\cite{Pimpinelli} in which surface adatoms approaching a
descending step are more likely to descend over the step rather than
being reflected from it, due to a step edge potential barrier.  This
creates a downhill flux of adatoms ($\mathbf{j} \propto -\nabla h$)
and a smoothing term ($\partial h/\partial t \propto - \nabla \cdot
\mathbf{j} $) identical to the first term in the KPZ
equation~\cite{Pimpinelli}. In practice, the atomic scale dynamics is
complex with surface reconstructions, complicated step edge
geometries, and a two component (Ga, As)
surface~\cite{Tejedor,Tersoff,Heyn}.  The sign of the ES effect in
GaAs is controversial~\cite{Ballestad,Coluci,Johnson}, but we show
below that a negative ES effect, favoring downhill flow (stable
growth) is consistent with the experimental data.

The nonlinear term in KPZ is associated with growth along the outward
normal, as in chemical vapor deposition.  In this case, $\lambda$
should be equal to the growth rate $F$.  However, the value for
$\lambda$ needed to simulate the experimental results is almost two
orders of magnitude larger than $F$~\cite{Ballestad}.  Also, the KPZ
nonlinearity is non-conservative, whereas MBE growth is conservative
with a growth rate that is independent of the surface shape.
\begin{figure}[tbp]
  \centering 
  \includegraphics[scale=1.3]{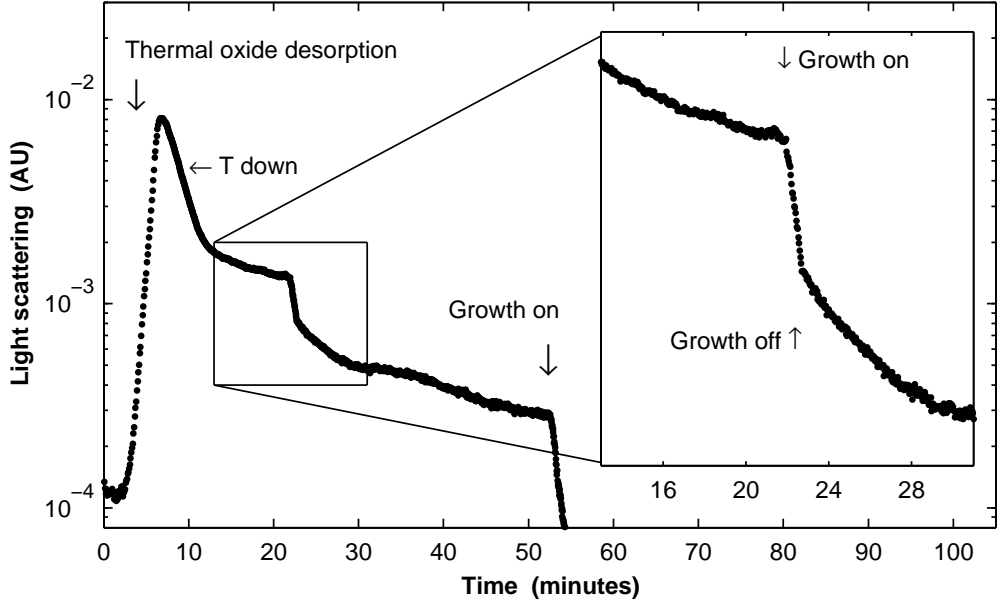}
  \caption{Light scattering during growth corresponding to 
    surface power spectral density at 41\,$\mu$m$^{-1}$ showing the
    effect of atom deposition on the smoothing rate.  The sample
    roughens during a temperature ramp to remove the surface oxide at
    about 5 minutes in the figure, which is followed by relatively
    fast smoothing during a high temperature (620$^{\circ}$C) anneal
    for about 7 minutes, and then slower smoothing during annealing at
    growth temperature (550$^{\circ}$C).  }
  \label{LS}
\end{figure}

\begin{figure}[tbp]
  \centering 
  \includegraphics[scale=1.3]{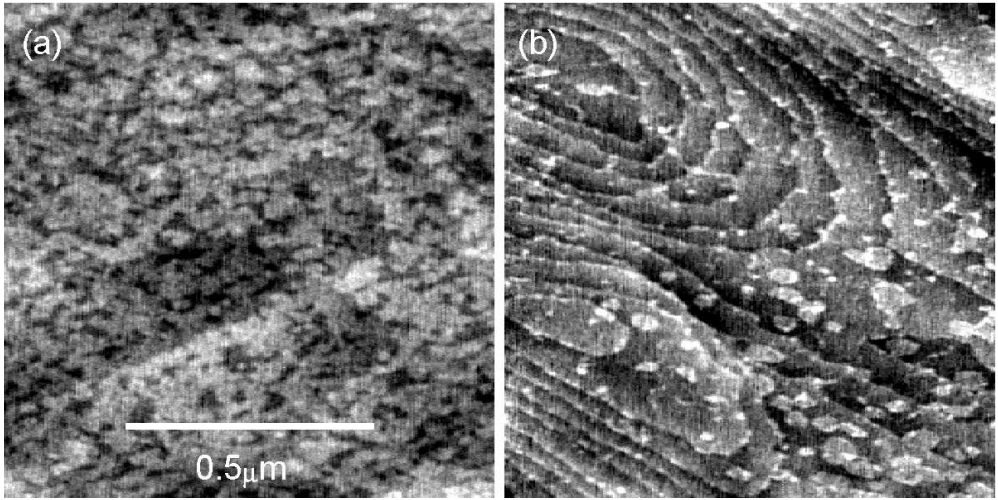}
  \caption{AFM images of (a)~a sample quenched 
    (fast cooled) after 69 minutes of growth at 600$^{\circ}$C and
    (b)~a sample annealed for 15 minutes at growth temperature
    595$^{\circ}$C after 40 minutes of growth.  }
  \label{AFM}
\end{figure}

In addition, the KPZ description with constant coefficients is not
consistent with experiments which show that the smoothing rate depends
on the growth rate. For example, in Fig.~\ref{LS} we show the
scattered light intensity from a GaAs surface during an interruption
in growth on a randomly textured substrate.  The intensity of
scattered light is proportional to the power spectral density of the
surface topography at a spatial frequency $q$ determined by
geometrical factors~\cite{Ballestad} (in this case
$q$=41\,$\mu$m$^{-1}$, corresponding to a lateral surface length-scale
of about 150\,nm).  For low amplitude surface textures, in the KPZ
model the surface should smooth exponentially with a characteristic
rate given by $\nu q^2$ where $q$ is the spatial frequency of the
surface roughness~\cite{Pimpinelli}.  As shown in the inset of
Fig.~\ref{LS}, the smoothing rate responds immediately to changes in
the growth flux; it is faster during deposition and slower during
annealing, suggesting that $\nu$ is flux dependent.  This continued
smoothing of the surface in the absence of an atom flux indicates that
the physical mechanisms at play on the surface still favor a net
downhill migration of surface adatoms, even after the flux of atoms
from the vapor has been turned off.

Insight into why the smoothing rate depends on the flux can be
obtained by comparing an atomic force microscope (AFM) image from a
sample which is fast cooled (quenched) after growth with the AFM image
of a surface which has been annealed (see Figs.~\ref{AFM}a and b).  The
quenched sample (a) is covered with small islands, whereas the
annealed sample (b) has broad terraces with few islands. The small
islands must coalesce into the step edges during annealing.  The
kinetic barrier to the adatom coalescence into the step edges, causes
the growth process to be non-local in space and time, in contrast to
KPZ.  A high density of steps at one location that absorb adatoms will
affect the adatom density and hence the growth rate at another nearby
location.

\section{Coupled Growth Equation Model}
\label{coupled}

The growth phenomena discussed above can be explained in a natural way
if we extend the growth model to include the adatom dynamics
explicitly with two coupled growth equations (CGE)~\cite{EarlierCGE}:
 \begin{subequations}
  \begin{eqnarray}
    \label{BTSn}
    \frac{\partial n}{\partial t} + \nabla \cdot \mathbf{j} &=& F -
    \frac{\partial h}{\partial t}, \\
    \label{BTSh}
    \frac{\partial h}{\partial t} &=& 2Dn^2+(\beta Dn-\kappa) S.
  \end{eqnarray}
  \label{BTSeqns}
\end{subequations}
\noindent Eqn.~\ref{BTSn} is a continuity equation for the adatom 
density $n$ with source and sink terms, while Eqn.~\ref{BTSh}
describes the time dependence of the surface height $h$, which depends
on the dimer nucleation rate and the net adatom attachment rate at
steps.  The constants are defined in atomic units as follows: $F$ -
deposition rate from the vapor, $D$ - adatom diffusion coefficient,
$S$ - density of steps, $\kappa$ - rate of thermal evaporation of
atoms from step edges into the adatom phase, and $0<\beta<1$ is the
sticking coefficient for an adatom at a step edge.  An adatom is
defined as a diffusing unit on the surface, which could be a Ga atom
or a Ga-As complex.  We also define:
\begin{subequations}
  \begin{eqnarray}
    \label{Jeqn}
    \mathbf{j} &=& -D(\zeta n \nabla h + \nabla n), \\
    \label{Seqn}
    S &=& \sqrt{S_0^2+(\nabla h)^2},
  \end{eqnarray}
  \label{SUPPeqns}
\end{subequations}
\noindent where in Eqn.~\ref{Jeqn}, $\mathbf{j}$ is the surface 
current of adatoms and $0<\zeta<1$ is a proportionality constant which
describes the net downhill drift of adatoms.  The second term in
Eqn.~\ref{Jeqn} represents adatom diffusion.  In Eqn.~\ref{BTSh}, any
adatom that attaches to a step edge is assumed to have incorporated
into the film.  The downhill drift parameter $\zeta$ can be positive
or negative: a positive value favors downhill drift of adatoms,
consistent with the surface smoothing that is observed experimentally
for GaAs $(001)$~\cite{Coluci,Ballestad} (and also consistent with a
negative value for the ES energy barrier).  

In Equation~\ref{Seqn}, we present a physically plausible hypothesis
for the dependence of the rms step density on the surface slope.  In
this expression the random local surface slope associated with the
growth-induced step density $S_0$ is added to the deterministic
macroscopic surface slope $\nabla h$.  Since the local slope
associated with the background step density $S_0$ is random, the two
terms add in quadrature.  In Equation~\ref{Seqn}, we assume that the
background step density is independent of the macroscopic surface
slope. We expect $S_0$ to depend on temperature and deposition rate,
and on time in the case of growth interrupts (see Fig.~\ref{AFM}a and
b)~\cite{Braun}.  The simple picture of a surface consisting of flat
terraces separated by atomic steps, can be expected to provide a good
description as long as the surface slope does not reach the next low
index crystal planes, namely (110) and (111).  These planes are
$45^{\circ}$ and $54.7^{\circ}$ from the surface normal, and beyond
the range of surface slopes that we have explored experimentally
($\lesssim 30^{\circ}$). We assume that the density of random steps
$S_0$ is independent of the topography. In this approximation, the
average step density is proportional to the rms value of the local
surface slope.  The expression for $S$ is then constructed by
averaging over the random orientation of the local slope, and the rms
step density is given by the incoherent sum of the two contributing
factors.

For low amplitudes and long wavelength ($\nabla h<S_0$), the adatom
density will be nearly constant as a function of position and time,
and approximately equal to $n_0=(F+\kappa S_0)/\beta DS_0$.  In this
case, Eqns.~\ref{BTSn},~\ref{BTSh} reduce to,
\begin{equation}
  \frac{\partial h}{\partial t} = \frac{\zeta}{\beta} 
    \left(\frac{F}{S_0}+\kappa\right)\nabla^2h+F.
  \label{BTSlin}
\end{equation}
\noindent This reproduces the linear part of the KPZ equation and 
shows explicitly the dependence of the linear smoothing coefficient
$\nu$ on the deposition rate and the downhill drift parameter $\zeta$.
In addition, it shows that in the absence of growth ($F=0$) the linear
smoothing term is independent of the background step density $S_0$.
This agrees with the light scattering data in Fig.~\ref{LS}, which
shows that the smoothing rate is relatively constant during a growth
interruption even though the AFM images in Fig.~\ref{AFM} indicate
that the step density drops dramatically during annealing.  Extending
Eqn.~\ref{BTSlin} to higher order, one finds non-linear terms with
higher order spatial derivatives. We speculate that the higher order
nonlinear terms can be approximated by the KPZ nonlinearity over a
limited spatial frequency range if the surface topography is not too
large.
\begin{figure}[tbp]
  \centering 
  \includegraphics[scale=1.3]{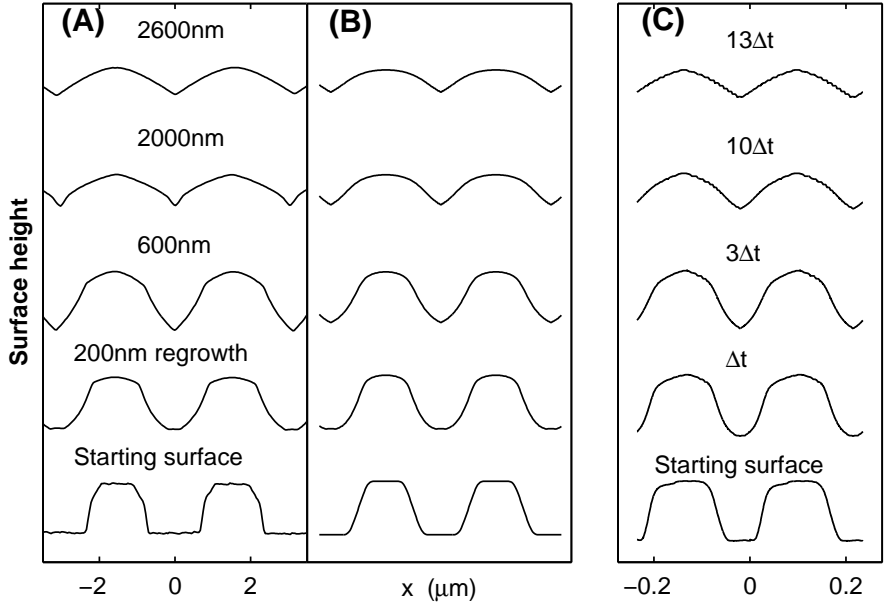}
  \caption{Film thickness dependence: (A) AFM scan lines for regrowth 
    on 100 nm deep gratings oriented perpendicular to the [$110$]
    direction; (B) Scan lines from CGE calculation; (C) Scan lines
    from 2D kMC simulation of 10 nm high grating structure, where one
    $\Delta$t equals $5.6$ ML of growth.  All offsets arbitrary. }
  \label{JENSgratings}
\end{figure}

\section{Textured Surfaces: Film Thickness Evolution}
\label{kmc}

Growth on substrates with larger amplitude surface slopes, up to
$\sim30^{\circ}$, show complex surface shapes before evolving into
parabolic mounds, as shown in Fig.~\ref{JENSgratings}a.  At
intermediate times the valleys are V-shaped with concave rather than
convex sidewalls and distinct shoulders near the top of the sidewalls.
Note the absence of $(100)$ facets which are predicted theoretically
for annealing below the roughening temperature in the absence of
deposition~\cite{Shenoy}.  Equations~\ref{BTSeqns} and~\ref{SUPPeqns}
can be solved in seconds with a finite difference scheme and a coupled
differential-algebraic system solver, and a 1D solution is shown in
Fig.~\ref{JENSgratings}b with parameters adjusted to match the
experimental data in Fig.~\ref{JENSgratings}a (see
Table~\ref{parameter_table} for parameters).  The agreement with the
experimental surface shapes is striking. In particular, the model
reproduces the inverted "Gothic window" shape of the valley for the
600 nm growth and the KPZ-like cusps in the 2600 nm growth where the
grating amplitude has reduced sufficiently so that the structure is
described by the KPZ equation.

A continuum model cannot include the microscopic details of the atomic
scale phenomena, such as the geometry and density of step edges.  We
therefore compare the continuum model in Eqns.~\ref{BTSeqns}
and~\ref{SUPPeqns} with a kinetic Monte Carlo (kMC) simulation, which
includes the same physical processes that are included in the
Eqns.~\ref{BTSeqns}.  We use a 2D, cubic grid, one-component,
restricted solid-on-solid (SOS) model, with nearest-neighbor
interaction.  Each atom bonds to the surface with an activation energy
$E_{act}=E_{sub}+mE_{lat}$, where $m$ is the number of lateral
neighbors~\cite{KMCexplained}.  The kMC simulations produce a random
step distribution automatically due to the statistical nature of the
model.  In kMC, the binding energy for an atom at a step edge depends
on how many neighbors it has ($\sim mE_{lat}$), whereas in the CGE
continuum model a single average value is used for the step edge
binding energy.

\begin{table}[htbp]
  \caption{Parameter table for CGE calculations.  Atomic units were 
    used with a lattice constant of 0.3 nm was used.}
  \label{parameter_table}
  \vspace{0.25cm}
  \begin{tabular}{cccccccc}  \hline
    Figure & T & \text{F} & D & \text{$\kappa$} & \text{S$_0$} &
    \text{$\beta$} & \text{$\zeta$} \\ \hline
    2 (b) & 580 & 1.0 & 180 & 3.0 & 0.075 & 0.3 & 0.075 \\ \hline
    3 (b) & 420 & 0.8 & 0.2 & 0.00025 & 0.025 & 0.1 & 0.15 \\ \cline{2-8}
    & 500 & 0.8 & 9.0 & 0.019 & 0.02 & 0.1 & 0.15 \\ \cline{2-8}
    & 550 & 0.8 & 60 & 0.19 & 0.02 & 0.2 & 0.15 \\ \cline{2-8}
    & 610 & 0.8 & 460 & 2.0 & 0.01 & 0.4 & 0.15 \\ \hline  
  \end{tabular}
\end{table}
SOS simulations of MBE growth by kMC are limited by available
computing power to small scale structures, and become intractable for
realistic, high temperature growth scenarios where 2D systems have
sides up to microns and growth times on the order of hours. In
Fig.~\ref{JENSgratings}c, we show a kMC simulation for a surface
grating that is somewhat smaller than the experimental structures.
The simulated grating profiles in Fig.~\ref{JENSgratings}c were
obtained by projecting 2D kMC simulations onto a line at each time
point by taking the average elevation perpendicular to the scan line.
In this simulation, $E_{sub}$=1.25~eV, $E_{lat}$=0.35~eV and an ES
step-edge barrier of $E_{ES}$=-0.05~eV was used for the downhill drift
mechanism.  The agreement with the experimental shapes is excellent,
reproducing all of the main features, except they are on a smaller
size scale. The substrate and lateral binding energies are similar to
values reported earlier in the interpretation of RHEED
data~\cite{BAJoyce,Shitara} and compatible with the fitting parameters
found in the continuum model.  It is plausible that similar shapes
could be obtained for the larger size scales relevant to the
experiments by scaling the parameters appropriately.  In the case of
the CGE model (Eqns.~\ref{BTSeqns},~\ref{SUPPeqns}) we find that the
parameters can indeed be scaled to yield similar surface shapes at
different length scales~\cite{ABphd}.

\section{Textured Surfaces: Temperature Evolution}
\label{comparison}

\begin{figure}[tbp]
  \centering \includegraphics[scale=1.3]{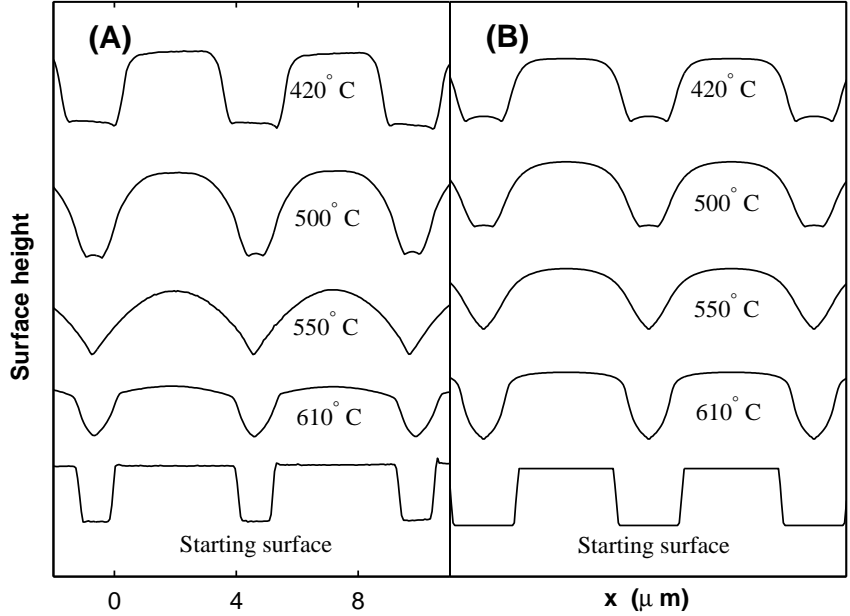}
  \caption{Growth temperature dependence: (A) AFM scan lines for regrowth 
    on 100 nm deep gratings oriented perpendicular to the
    [$1\bar{1}0$] direction; (B) CGE calculation.  The grating pitch
    is $5\mu$m.  All growths are 1 hour at $0.8$ML/s.  All offsets
    arbitrary.}
\label{t_depn_fig}
\end{figure}

In Fig.~\ref{t_depn_fig}a, we show the dependence of the surface
topography on growth temperature, for a fixed layer thickness together
with (b) the simulated surface topography using Eqns.~\ref{BTSeqns}
and~\ref{SUPPeqns} and parameters as outlined in
Table~\ref{parameter_table}. The experimental data is obtained from
growths on 100 nm deep gratings oriented perpendicular to
$[1\bar{1}0]$.  This is the faster diffusion direction in this
material system~\cite{Ballestad}, and depends on the As$_2$/Ga ratio
during growth, which was equal to three for all growths discussed in
this work.  This observation is consistent with the values used for
the downhill drift parameter in our calculations, where the best fits
were obtained using a larger $\zeta$ when the gratings were oriented
perpendicular to $[1\bar{1}0]$ (Fig.~\ref{t_depn_fig}b) than for
gratings perpendicular to the $[110]$ direction
(Fig.~\ref{JENSgratings}b).  The diffusion constant D, however, was
considered isotropic in all calculations in this paper.  There is some
uncontrolled variation in the pitch and depth of the gratings in the
experimental data in Fig.~\ref{t_depn_fig}a.  Nevertheless, the CGE
model reproduces the main features in the temperature dependent data,
namely the small secondary mound in the valley at 500$^\circ$C, the
KPZ-like cusp at 550$^\circ$C and the inverse Gothic window shape for
the valleys at 610$^\circ$C.  The shoulders at the edges of the ridges
at 610$^\circ$C are also reproduced by the model, although they are
not as sharp as in the experimental data.

The parameters used in these calculations are based on the same
energies used in the kMC simulations in Fig.~\ref{JENSgratings}c.  The
diffusion constant is related to the substrate binding energy through
the expression $D=(2kT/h) \exp{(-E_{sub}/kT)}$. The step edge
detachment rate is calculated from $\kappa=D
\exp{(-\hat{m}E_{lat}/kT)}$, where $\hat{m}$ is an average number of
neighbors for atoms at a step edge which we set equal to 2.25.  The
declining value of $S_0$ with temperature is reasonable; one might
also expect $\zeta$ to decrease weakly with temperature.  Satisfactory
fits to the data were also obtained with a larger activation energy
for $D$ (1.8~eV rather than 1.25~eV) together with a smaller prefactor
and somewhat different (yet still physically reasonable) temperature
dependences for the other parameters. Experimental and theoretical
work suggests that the activation energy for $D$ is in the 1.5-2.0~eV
range.~\cite{BAJoyce,Shitara,Thibado,Kratzer}.

\section{Conclusions}
\label{conclusion}

We have shown that the complex surface morphology which develops
during epitaxial regrowth on patterned GaAs $(100)$ substrates, can be
explained by the dynamics of the deposited adatoms, including step
edge attachment and detachment, adatom diffusion, and downhill drift.
Although we attribute the downhill drift to a negative
Ehrlich-Schwoebel barrier we cannot rule out the possibility that this
effect is caused by some other mechanism.  This analysis is specific
to GaAs, but the concepts are generic and may be applicable to other
systems.




\begin{thebibliography}{00}
\bibitem{EarlierCGE}Related coupled equations have been used earlier
  to describe growth on faceted surfaces~\cite{Koshiba,Ploog}. In this
  work the adatom incorporation rate was dependent on the crystal
  facet orientation.
\end{thebibliography}
\end{document}